# SINFONI in the Galactic Center: young stars and IR flares in the central light month[1]

F.Eisenhauer[1], R.Genzel[1,2], T.Alexander[3,6], R.Abuter[1], T.Paumard[1], T.Ott[1], A.Gilbert[1], S.Gillessen[1], M.Horrobin[1], S.Trippe[1], H.Bonnet[4], C.Dumas[4], N.Hubin[4], A.Kaufer[4], M.Kissler-Patig[4], G.Monnet[4], S.Ströbele[4], T.Szeifert[4], A.Eckart[5], R.Schödel[5] & S.Zucker[3]

[1] Max-Planck Institut für extraterrestrische Physik (MPE), Garching, FRG

[2] Department of Physics, University of California, Berkeley, USA

[3] Center for Astrophysics, Weizmann Institute of Science, Rehovot, Israel

[4] European Southern Observatory (ESO), Garching, FRG

[5] 1.Physikalisches Institut der Universität Köln, FRG

[6] Incumbent of the William Z. and Eda Bess Novick career development chair

## Abstract

We report 75 milli-arcsec resolution, near-IR imaging spectroscopy within the central 30 light days of the Galactic Center, taken with the new adaptive optics assisted, integral field spectrometer SINFONI on the ESO-VLT. To a limiting magnitude of K~16, 9 of 10

[1] based on observations obtained at the Very Large Telescope (VLT) of the European Southern Observatory, Chile

stars in the central 0.4”, and 13 of 17 stars out to 0.7” from the central black hole have spectral properties of B0-B9, main sequence stars. Based on the 2.1127μm HeI line width all brighter early type stars have normal rotation velocities, similar to solar neighborhood stars.

We combine the new radial velocities with SHARP/NACO astrometry to derive improved 3$^d$ stellar orbits for six of these ‘S’-stars in the central 0.5”. Their orientations in space appear random. Their orbital planes are not co-aligned with those of the two disks of massive young stars 1-10” from SgrA*. We can thus exclude the hypothesis that the S-stars as a group inhabit the inner regions of these disks. They also cannot have been located/formed in these disks and then migrated inwards within their planes. From the combination of their normal rotation and random orbital orientations we conclude that the S-stars were most likely brought into the central light month by strong individual scattering events.

The updated estimate of distance to the Galactic center from the S2 orbit fit is $R_o = 7.62 \pm 0.32$ kpc, resulting in a central mass value of $3.61 \pm 0.32 \times 10^6 M_\odot$.

We happened to catch two smaller flaring events from SgrA* during our spectral observations. The 1.7-2.45μm spectral energy distributions of these flares are fit by a featureless, ‘red’ power law of spectral index $\alpha$’=-4±1 ($S_\nu \sim \nu^{\alpha'}$). The observed spectral slope is in good agreement with synchrotron models in which the infrared emission comes from accelerated non-thermal, high energy electrons in a radiative inefficient accretion flow in the central R~10 $R_s$ region.

# 1. Introduction

Because of its proximity the Center of the Milky Way is a unique laboratory for studying physical processes that are thought to occur generally in galactic nuclei. The central parsec of our Galaxy contains a dense star cluster, with a remarkable number of luminous and young, massive stars (Forrest et al. 1987, Allen, Hyland & Hillier 1990, Krabbe et al. 1995, Genzel et al. 2000, Paumard et al. 2001), as well as several components of neutral, ionized and extremely hot gas (Mezger et al. 1996, Morris & Serabyn 1996, Genzel 2001). This constitutes a 'paradox of youth' (Ghez et al. 2003). For two decades, evidence has been mounting that the Galactic Center harbors a concentration of dark mass associated with the compact radio source SgrA* (diameter about 10 light minutes, Bower et al. 2004), located at the center of that cluster. Measurements of stellar velocities and (partial) orbits have established a compelling case that this dark mass concentration is a massive black hole of about 3 to $4x10^6$ $M_\odot$ (Eckart & Genzel 1996, 1997, Ghez et al. 1998, Schödel et al. 2002, 2003, Ghez et al. 2003, Ghez et al. 2005). The Galactic Center thus presently constitutes the best evidence we have for the existence of massive black holes in galactic nuclei. The Galactic nucleus provides a unique opportunity for studying environment and physical/stellar processes in the vicinity of a massive black hole.

Adaptive optics (AO) assisted, infrared integral field spectroscopy at the diffraction limit of 8m class telescopes promises valuable new information on the properties, kinematics and evolution of the nuclear star cluster down to main sequence masses of 4-8 $M_\odot$ (K~17-16). From the much increased sample of stars with both proper motion and radial velocity data it will be possible to significantly improve our knowledge of the stellar dynamics and mass distribution in the central region and determine the Galactic center distance to a precision of better than a few percent (e.g. Eisenhauer et al. 2003, Weinberg, Milosavljevic & Ghez 2004). Integral field spectroscopy will also explore the evidence for unusual stars that are predicted by theoretical models to exist in the dense environment near the black hole (Alexander 2003). Finally, time resolved measurements of the spectral properties of time variable emission from SgrA* itself will give new information on the innermost accretion zone, just outside of the event horizon.

In the present paper we report first observations of the Galactic Center with SINFONI, the first infrared AO integral spectrometer now operational on an 8m class telescope (Eisenhauer et al. 2003b,c, Bonnet et al. 2003, 2004). Our measurements shed new light on stellar properties and orbits in the central light month and on the nature of infrared variable emission from SgrA*.

## 2. Observations

The data were taken in July and August 2004 during the commissioning and first guaranteed time observations of SINFONI (Bonnet et al. 2004) at the Cassegrain focus of VLT-UT4 (Yepun). SINFONI consists of a cryogenic integral field spectrometer with a $2k^2$ Hawaii detector (SPIFFI, Eisenhauer et al. 2003b,c), coupled to a curvature sensor, adaptive optics unit (MACAO, Bonnet et al. 2003). We employed the highest resolution mode with 0.0125"x 0.025" (EW x NS) pixels fully covering a contiguous field of view of 0.8"x0.8". On July 15th we used the H+K grating simultaneously delivering for each spatial pixel ~2000 spectral pixels of FWHM 7-8x$10^{-4}$ μm (resolving power 2000-3400) across the entire 1.45 to 2.45μm region. On July 17th and August 18/19th we used the K-band grating delivering a FWHM resolution of 5x$10^{-4}$ μm (resolving power 4400). We locked the 60 actuator MACAO AO system on a 14.6 mag visible star ~20" NE of SgrA*. For that separation, the ~0.5-0.6" seeing and the ~3ms coherence time (in the visible) prevailing during the observations and a closed-loop bandwidth of ~60 Hz, the resulting Strehl ratio toward SgrA* varied between ~6 and 10% between the short end of the H and the long wavelength of the K-band. The stellar images in reconstructed cubes had 75 mas FWHM diameter, close to the diffraction limit of the VLT in the K-band (~60 mas).

On July 15th (02:37 – 05:15 UT) we acquired 11 individual data cubes of 10 minutes each, spatially dithered to cover a region of ~0.9"x0.9" centered on SgrA* at an effective sampling of 12.5mas. On July17th (03:36 – 06:24 UT) we acquired ten data cubes of ten minutes each over about the same region. On August 18/19th (23:20-01:36) we covered this central region again with eight exposures of ten minutes each, the effective integration times varying from 20 to 80 minutes across the field. After dark subtraction, flat-fielding, wavelength calibration and dead/hot pixel interpolation, the final data cubes were divided by the spectrum of an A0V star to compensate for atmospheric absorption. We eliminated OH sky emission and background by observing several times throughout the observing a relatively star-free, ‘off’-field north-east of the Galactic Center, and then by subtracting the off-cube from the source data.

# 3. Results

Figure 1 shows the SINFONI ‘K-band’ image in August, in comparison with a FWHM 40mas resolution NACO (Lenzen et al. 1998, Brandner et al. 2002) H-band image taken about two months prior. The SINFONI images were constructed by collapsing the data cubes between 2 and 2.3μm along the spectral dimension. The FWHM resolution of the imaging spectroscopy is 75mas and all NACO stars <16.8 mag are clearly detected in the SINFONI image. Our data thus give new information on the spectral properties of the ‘S-stars’ in the immediate vicinity of the black hole (Eckart & Genzel 1997, Ghez et al. 1998, Schödel et al. 2003, Ghez et al. 2005). In two data sets (July15$^{th}$ and 17$^{th}$) we also were lucky to catch flaring activity from SgrA*, thus giving the first direct information on its spectral properties.

## 3.1 Most of the ‘S-stars’ are B main-sequence stars

Figures 2, 3 and 4 show background subtracted spectra near the 2.1661μm HI Brγ line for 21 out of the 25 K<16 mag stars[2] in the central 0.7” radius region, plus spectra toward

[2] Two of the 21 stars at the SINFONI resolution (S17 and id26) are actually double, as demonstrated by the higher resolution NACO images of Fig.1. S17 appears to consist of one early and one late type star separated by ~0.07” NE-SW (spectra in Figs. 2-4) and id26 consists of two early type stars (Fig.4) separated by ~0.1” EW. For four additional stars at the edge of the SINFONI field of view (S6, id23, S11

SgrA* itself. By subtracting background apertures close to the stellar sources most of the strong foreground/background nebular interstellar Brγ emission centered near the local standard of rest (LSR) velocity (and stretching between –150 and +250 km/s) is eliminated. We then excised any small residual interstellar emission component by interpolation over it in each spectrum. For the brightest stars we also show spectra near the 2.1127μm HeI line. Interstellar emission is negligible for this line.

Of the 10 K≤16 stars in the central 0.4” 9 exhibit prominent Brγ absorption lines characteristic of the infrared properties of early type stars (Hanson, Conti & Rieke 1996). The brighter stars exhibit HeI in absorption as well. Over the central 0.7” radius region there are 14 to15 early and 9 late type stars. If the magnitude of proper motion or total velocity is used in addition to screen against stars in the fore- and background that are merely projected into the central region, the early star fraction increases over the entire field. Taking a total (3d) velocity of 500 km/s as the dividing line, appropriate for circular motion at R=0.04pc around a central mass of $4x10^6$ M$_\odot$, there are 10 early type stars and ***no*** late type stars with K≤16 that are plausibly located within the central 0.7”. Earlier evidence showed the lack of late type absorption features toward the S-cluster region (Genzel et al. 1997, Eckart, Ott & Genzel 1999, Figer et al. 2000, Gezari et al. 2002). S2 was the first star for which Ghez et al. (2003) unambiguously proved its nature as an O9/B0 main sequence star. Our data now show that arguably ***all*** brighter stars near the black hole are early type stars, thus increasing yet further the ‘paradox of youth’ (Ghez et al. 2003). How have these stars come to reside so close to the central black hole?

---

and W6) we do not show spectra. Three are late type stars. S6 may also be early type but the SNR ratio on our possible Brγ detection is marginal.

Our new spectra permit us to derive radial velocities (Table 1) for all 15 Brγ absorption line stars (Figures 2, 3, 4). In addition we see indications for two or three absorption features (–1900, -1100 and +1600 km/s in July and -1700 and -1000 in August) toward SgrA* itself. If real these features originate most likely in stars along the line of sight toward and behind SgrA*. The features at –1100 km/s and –1900 km/s may be due to the spillover of light from S2 and the double source S17, respectively. The feature at +1600 km/s may come from another, fainter star closer to the position of SgrA*, perhaps also visible in the higher resolution NACO H-band image just south of SgrA*. Not unexpectedly, this does indicate that the 'quiescent' emission toward SgrA* has a stellar contribution (Genzel et al. 2003b).

The brighter (K<15) mag stars all exhibit FWHM ~400 to 600 km/s Brγ absorption lines with a Lorentzian profile. The 2.1127μm HeI line is significantly narrower. The fainter stars (K~15-15.8) exhibit significantly broader Brγ absorption (FWHM ~500 to 1700 km/s) and not much HeI absorption. This difference becomes especially clear if we further enhance the quality of the spectra by co-adding the spectra of S-stars, split into a brighter (K~14-15: S1, S2, S4, S8, S9) and a fainter (K~15-15.8: S5, S08, S12, S13, S14) group. The co-added spectra, obtained by shifting the individual spectra to the wavelength of maximum absorption are shown in Figure 5. Excess Brγ absorption/emission redward of the minimum in some of the blue stars (e.g. S1 and S2) as well as in the co-added spectrum of the brighter star group may be a combination of poorly eliminated interstellar emission in the range -500 to 500 km/s LSR and spillover from other sources in the crowded central region.

For the brighter group, the average spectrum exhibits strong absorption lines at 2.1661 μm (Brγ) and 2.1127μm (HeI). In the H-band we marginally detect two further hydrogen Bracket lines (10-4, 11-4), plus the 1.701μm HeI line. The equivalent widths of Brγ and 2.1227μm HeI are 5.8(±0.4) Å and 1.3(±0.3) Å. Average FWHM line widths are 460 (±18) km/s for Brγ and 290 (±35) km/s for HeI. For the fainter group we only detect Brγ with an average equivalent width of 11.7 (±2.5) Å and a FWHM of 1200 (±80) km/s. We set a $2\sigma$ upper limit to the HeI equivalent width of 1 Å. A comparison to the infrared spectral atlas of Hanson, Conti & Rieke (1996) and Hanson et al. (2004) shows that these spectral characteristics, including also the difference in line width and shape between Brγ and HeI, are very well matched by B main sequence stars. In Figure 6 we compare the deduced equivalent widths to the values of solar neighborhood stars of different spectral type and luminosity class (Hanson et al. 1996). For the brighter group the combination of Brγ and HeI equivalent widths are matched by B0-B2 V stars. The fainter group has properties of B4-9 V stars. There is no match to spectra of giants and supergiants. These spectroscopic identifications are in very good agreement with their photometry (<K>~14.5 for the brighter group and ~15.5 for the fainter group) if a K-band extinction of 2.8mag is adopted from a fit to the S-star colors. This extinction value is derived from the S-star colors for a ~35,000 K star and is uncertain by ±0.2 mag. The relative ratios of the marginally detected HI 10-4 and 11-4 and 1.7μm HeI lines is also consistent with early BV stars in the H-band atlas of Hanson, Rieke & Luhman (1998).

The ratio in line width of Br$\gamma$ to 2.1127μm HeI of ~1.6 is consistent with solar neighborhood B stars as well. Unlike the pressure broadened Br$\gamma$ line, the 2.1127μm HeI has an intrinsic line width of 280 km/s FWHM (taking into account the ~80 km/s instrumental profile). This sets a limit to the rotational velocity of the S-stars of $<v_{rot}\sin(i)>$~0.55 FWHM = 154(±19) km/s. For comparison, Gathier, Lamers & Snow (1981) find $<v_{rot}\sin i>$~130 km/s for a sample of solar neighborhood, early B-stars. The increase in Br$\gamma$ line width for later B stars (and A-stars) , owing to the larger neutral hydrogen abundance at the bottom of the photosphere in these stars, is also seen in the data of Hanson et al. (2004). Note that in spite of the high collision rate in the stellar cusp, stochastic tidal spin-up (Alexander & Kumar 2001) is not expected to significantly affect the relatively short-lived B-stars.

In summary the overwhelming evidence from our new SINFONI spectra is that the $K<16$ mag S-stars within 30 light days of the black hole are normally rotating, main sequence B-stars.

### 3.2 Global fit to stellar orbits in the central 0.5”

We have used the new radial velocity measurements, in conjunction with our 2004 NACO data and the 1992-2003 SHARP/NACO data of Schödel et al. (2003) and Eisenhauer et al. (2003a) to obtain improved estimates of the orbital parameters of the S-stars. Proceeding as in Eisenhauer et al. (2003a) and in Schödel et al. (2003), we fitted bound Kepler orbits to the 1992-2004 spatial and velocity data of S1, S2, S4, S5, S6, S7, S8, S08, S9, S12, S13, S14, S17 and id26e. For six stars (S1, S2, S8, S12, S13, S14) our position and radial velocity data allow the proper determination of all orbital elements. We then fitted simultaneously the orbits of these six stars treating the central mass and its location ($x_o$, $y_o$) as free fit parameters, and assuming that the gravitational potential is dominated by the central point mass (Schödel et al. 2003, Mouawad et al. 2004). If not specified otherwise (section 3.3) we adopt throughout this paper $R_o$=8 kpc for the Galactic Center distance (Reid 1993, Eisenhauer et al. 2003a). We also assumed that the intrinsic velocity of SgrA* is negligible, an assumption that appears well justified based on the recent VLBA proper motion measurements of SgrA* by Reid & Brunthaler (2004). Reid & Brunthaler find that the proper motion of SgrA* is less than about 20 km/s in the plane of the Milky Way, and 2 km/s (2σ) perpendicular to it. Ghez et al. (2005) deduce a limit of SgrA*'s proper motion of 60 km/s from their most recent analysis of the Keck IR data. Table 1 lists the orbital parameters for the global fit of the 6 S-stars for which we were able to deduce a complete and well determined set of orbit parameters (Fig. 8a). The best fit central mass is 4.06 (±0.38) $\times 10^6 M_\odot$ (for a distance to the Galactic center of 8kpc), where the quoted uncertainty is the 1σ total error, including the ±0.32 kpc uncertainty in the distance to the Galactic Center (see section 3.3). Because of the availability of radial velocities, our new analysis allows, for the first time, an unambiguous determination of the $3^d$-orientation of the best six orbits (Fig.8a). For an additional 8 stars we were able to constrain the orientation of their angular momentum vectors to a curve on the sky (Fig.8b). The orientations of the S-stars appear to be random in space, with respect to each other, as well as with respect to the two rings/disks of massive young stars at radii of 1-10" (Levin & Boloborodov 2003, Genzel et al. 2003a,

Paumard et al. 2005). This confirms and strengthens the conclusions of Ghez et al. (2005).

As will be discussed further in section 4.1, the orbital spin precession due to the Lense-Thirring frame dragging effect around a spinning black hole is a key parameter for the orbiting stars in the central light month (Levin & Beloborodov 2003). The last row in Table 2 lists the Lense-Thirring precession periods for the different stars, obtained from equation (5) of Levin & Boloborodov (2003) with spin parameter 0.52 +/- 0.15 (Genzel et al. 2003b). For S2 and S14 these precession time scales are comparable to the average stellar ages in the two outer star rings/disks. We thus conclude that S2 and S14 would have lost the 'memory' of their original orientation as the result of the Lense-Thirring precession (see also Levin & Boloborodov 2003) if they had originally been in the plane of these disks and were then subsequently scattered into the central region. Not so, however, for the other 4 stars in Fig.8a and Table 2 (S1, S8, S12 and S13) and for 6 of the 7 stars in Fig.8b whose Lense-Thirring precession periods are much greater than the age of the outer star disks. Only one of the four best determined orbits in Fig.8a with suitably long Lense-Thirring precession periods (S13) could plausibly lie in the plane of one of the (the counter-clockwise) outer young star disks (Levin & Beloborodov 2003, Genzel et al. 2003a). Two (S6, S7) to maximum five (S4, S5, S08) of the stars with more poorly determined orbital orientation could potentially lie in the planes of the outer star disks (Fig.8b). Three of the well-defined orbits (S1, S8, S12) and another three to six (S4, S5, S08, S9, S17, id26e) of the less well defined orbits are not coplanar with these disks. In total, 50-75% of the S-star orbits with long Lense-Thirring precession periods are not aligned with the outer disks.

In summary then of this section we can now firmly exclude that the S-stars - taken together as a group - are co-planar with the two outer star disks, either because the S-stars inhabit the innermost regions of these disks, or because they were once located or formed further out in the disks and gradually migrated inward within their planes.

### 3.3 Distance to the Galactic center

The additional radial velocity measurements and position data from 2004 allow us to update the geometric distance measurement to the Galactic center from Eisenhauer et al. (2003a). Eisenhauer et al. reported $R_o = 7.94 \pm 0.42$ kpc. Our new orbit fit to the available data of S2 treating the distance, position, and mass of the central black hole as a free parameter results in $R_o$= 7.62+/-0.33 kpc and a reduced chi square $\chi^2_{(r)} = 0.70$. Applying the error scaling to $\chi^2_{(r)} = 1$ and adding the systematic error of ±0.16 kpc from the unknown residual motion of our reference coordinate system as outlined in Eisenhauer et al. (2003a), our best estimate for the distance of the Galactic center now is $R_o = 7.62 \pm 0.32$ kpc. The resulting central mass is $3.61 \pm 0.32 \times 10^6\ M_\odot$. Inclusion of the other 5 stars with well defined orbits results in a similar distance value but does not improve the uncertainty.

## 3.4 Spectral energy distribution of infrared flares from SgrA*

Genzel et al. (2003b) and Ghez et al. (2004) reported the first detections of variable infrared emission from SgrA* itself. The source of infrared variable emission flares is coincident with SgrA* and the position of the dynamical center to an accuracy of a few milli-arcsecs. The infrared flares probably come from hot or relativistic gas within a few times the event horizon (Genzel et al. 2003b, Ghez et al. 2004, Yuan, Markoff & Falcke 2002, Yuan, Quataert & Narayan 2003, 2004, Liu, Petrosian & Melia 2004). The key for determining the emission processes is the measurement of the spectral slope and polarization of the variable infrared emission and the simultaneous observations of infrared and X-ray flares (Baganoff et al. 2001, Porquet et al. 2003, Goldwurm et al. 2003, Eckart et al. 2004). Theoretical models predict that the infrared emission comes from a small fraction (a few percent) of the electrons near the event horizon that are accelerated to $\gamma \geq 10^3$ in a non-thermal distribution and radiate in the near-infrared via synchrotron emission (Yuan et al. 2004). The non-simultaneous H, $K_s$, and L' band observations of 4 flares with the VLT in 2004 raised the tantalizing possibility that the spectral slope might be blue, inconsistent with the synchrotron model, and possibly requiring thermal emission from hot, optically thick gas (Genzel et al. 2003b).

During our SINFONI observations we were lucky to catch in full length two smaller flares of SgrA*. These flares were equivalent to a de-reddened K-magnitude of ~16, increasing by about one magnitude above the 'quiescent' state emission toward SgrA* (after correction for spillover of light from S2, S13 and S17). For comparison, the brighter flares seen by Genzel et al. (2003b) reached K~15. Both flares (on July 15$^{th}$ and July 17$^{th}$, 2004) occurred toward the position of the compact radio source, to within our measurement accuracy of ±12 mas (Figure 9, top inset). In each case we detected excess emission over about an hour. To determine the spectral energy distribution (SED) of these flares, we first divided the aperture source counts for each of the frames by the corresponding S2 source counts. This removes the effects of wavelength dependent Strehl ratio, atmospheric absorption and interstellar extinction. Next we subtracted from each frame the 'off'-flare SED to eliminate the influence of scattered light/spillover and foreground/background stellar emission. The left inset of Figure 9 shows the final SED average over three frames near the maximum of the July 15$^{th}$ flare (filled (red) circles and ±1σ errors) and averaged over three frames during the rise and decay of the flare (open (blue) circles). To increase signal to noise ratio we binned the data at $\lambda \geq 1.97\mu m$ over 21 spectral points to an effective spectral resolution of 0.0105μm, and below 1.8μm over 101 pixels to 0.05μm. Because of the combined effects of extinction, decreasing Strehl and intrinsic slope, flux from the flare is not significantly detected below about 1.7μm. We have thus only included measurements above this wavelength for the SED analysis below. K-band data on July17th (right inset of Fig.9) were binned over 20 pixels to a resolution of $4.9 \times 10^{-3}$ μm.

The resulting SEDs of both data sets are clearly 'red' and featureless to within our measurement uncertainties. The flux density $S_\nu$ and the luminosity per logarithmic frequency interval, $\nu L_\nu$, decrease with frequency (or increase with wavelength). A power law fit ($S_\nu \sim \nu^{\alpha'}$, $\nu L_\nu \sim \nu^{\alpha}$, $\alpha = \alpha' + 1$) gives a spectral slope at the peak of the flares of $\alpha = -2.2 \pm 0.3$ for July 15$^{th}$ and $\alpha = -3.5 \pm 0.4$ for July 17$^{th}$. In the rising and falling flanks of the July 15th flare the slope was α=-3.8±0.8. The quoted uncertainties are 1σ statistical fit errors. Including systematics and data analysis we conclude that all data are fit by a red power law slope of common slope α=-3±1 (α'=-4±1). We can place a 2σ upper limit of about $10^{-2}$ $L_\odot$ to the dereddened luminosity of any moderately broad Brγ emission (Δv~ a few thousand km/s) during the flare.

# 4. Discussion

## 4.1 Paradox of Youth

The results presented here confirm and substantially deepen the mystery surrounding the existence of the massive young stars in the inner few hundredths of a parsec around the central black hole of the Galaxy. In short, the problem is that according to standard scenarios of star formation and stellar dynamics the stars cannot be born in such an extreme environment because of the strong tidal shear, but are also too short-lived to have migrated there from farther out. None of the solutions proposed so far for the puzzle of the young stars (e.g. Morris 1993, Genzel et al. 2003a; Ghez et al. 2003, 2005 and references therein) are entirely satisfactory. These broadly fall into three categories: (1) exotic modes of star formation near the massive black hole; (2) rejuvenation of old stars in the local population; (3) accelerated dynamic migration, or capture of stars from farther out. The different models are briefly summarized in Alexander & Livio (2004). Here we focus on the implications of our new findings.

Any explanation of the S-stars has to account for three principal facts.

- The stellar properties. The stars appear to be *entirely normal* early type B0V to B9V stars, in terms of their (extinction corrected) luminosities, their absorption lines equivalent widths and line ratios. In particular, their rotational velocities are similar to those of nearby Galactic disk B-stars.
- The spatial concentration. The relative fraction of the young stars increases toward the center, to the near exclusion of any late type stars in the inner ~0.02 pc.
- The orbital properties. The stellar orbits appear overall random, in marked contrast to the ordered planar rotation observed for the much more luminous emission line stars farther out. In addition the stars in the central 0.02pc appear to have orbital apo-apses greater than 0.01 pc (Ghez et al. 2005) and higher than random eccentricity (Schödel et al. 2003) although the latter trend is only of modest statistical significance (2-3σ) because of the relatively small number of stars observed.

An important open question is the connection, if any, between the S-stars inside ~0.02 pc and the luminous emission line stars further out, on the 0.1-1 pc scale. While it is plausible to assume that these are different components of the same parent population it should be noted that the two groups have distinct locations, kinematics and stellar properties. As we have discussed above, the present observations exclude now with some certainty that the S-stars as a group are located in the planes of the outer stars. The S-stars also do not follow the largely circular motions of the outer young stars. The S-stars are B dwarfs (S2, the brightest, is a transitional O8V-B0V star; Ghez et al. 2003) while the stars in the two young star disks at p~1-10" detected so far are more luminous OB supergiants, giants and Wolf-Rayet (WR) stars of various kinds (Genzel et al. 2003a, Paumard et al. 2001, 2005). This distinction is important, since B stars are more numerous, less massive (by about an order of magnitude) and longer lived (by up to two orders of magnitude) than the luminous O/WR stars. This significantly relaxes the constraints on the models. Future SINFONI observations will have to demonstrate whether the main sequence B-star population also is present outside the central cusp. If the S-stars initially were formed or located in one or the other of the outer stars disks, their original orbits must have been strongly perturbed.

The most important new result is the apparent normalcy of the B-stars. At face value this would appear to argue against any process that involves strong perturbations of the star: e.g. mergers (Genzel et al. 2003a), tidal heating (Alexander & Morris 2003) or stripping (Hansen & Milosavljevic 2003). However, this is by no means a conclusive argument, because it is not clear how efficiently a star relaxes after a major perturbation, and in particular how angular momentum is lost from the star or redistributed in it.

A relevant comparison can be made with the properties of blue stragglers, whose formation is thought to involve mergers, binary coalescence or 3-body interactions (Bailyn 1995). In spite of their violent birth, observations show that in many cases, late-B blue stragglers in intermediate age globular clusters, and early-B and O blue stragglers in open clusters do not seem to rotate faster than normal stars of the same mass in the field. This is in spite of the fact that, unlike lower-mass stars (later than F5), OB-blue stragglers do not have efficient magnetic breaking (Leonard & Livio, 1995 and references therein). In this context it is interesting to note the intriguing claim of a circumstellar disk around a blue straggler (de Marco et al. 2004), which may provide a breaking mechanism via magnetic anchoring. Neither is it clear whether collisional merger products should be mixed, and consequently, whether they should display unusual photospheric abundances. Lombardi, Rasio & Shapiro (1995, 1996) find that the mixing by the collision itself is minimal (at least for lower mass stars near a globular cluster's turnoff-mass). Sandquist, Bolte & Hernquist (1997) find similar results. However, subsequent convection or meridional circulation can still induce mixing (Leonard & Livio 1995; Lombardi et al 1995), although this was not found to be the case in detailed numerical work by Sills et al. (1997) and Oullette & Pritchet (1998). In summary, ”absence of proof is not proof of absence”. If the analogy to blue stragglers is justified, then it appears that the lack of unusual spectral features in the S-stars does not place strong constraints on their origin.

Next we consider the possibility that the S-cluster stars were initially associated with the stellar disks. Since migration in the disk plane is ruled out by the observed orientations of the orbits, we hypothesize that the orbits were subsequently randomized. According to their stellar contents, the two stellar disks/rings are coeval to about 1 Myr, and only ~5 Myr old (Krabbe et al 1995, Genzel et al 2003, Paumard et al. 2005). If the B-stars are associated with the disks, then the relevant time constraint for redistributing the orbits is the age of the disks, and not the considerably longer main-sequence lifespan of a B-star. The yet unspecified randomization mechanism must also work selectively on the B-stars and not on the other, more massive ring members. Two-body scattering, in addition to not being selective, is also too slow to account for the randomization. The typical two-body relaxation time in the inner GC is ~$10^9$ yrs for a solar mass star (e.g. Alexander 2003). Even if that time scale is correspondingly shorter for the more massive stars, it is clearly longer than a few Myr for stars of all masses given the high velocity dispersion close to the massive black hole. Alternatively, Milosavljevic & Loeb 2004 suggested that a star may be ejected to an S-star like orbit by a nearly grazing, 3- or 4-body interaction with other stars in its ``birth fragment'' in the disks. However, such collisions are rare and it is not known whether their rate can be high enough.

Orbital node precession, either due to general relativistic effects in the gravitational potential of the black hole (pro-grade), or due to Newtonian physics for a gravitational potential deviating from that of a point mass (retro-grade), are both important on the time scale of a few Myrs (Jaroszynski 1998, Fragile & Matthews 2000, Rubilar & Eckart 2001) but do not change the orientation of the angular momentum vector of the orbit, as long as the gravitational potential (e.g. of the extended central stellar cusp) is spherically symmetric. They change, however, the orientation of the apo-apse (see discussion in Ghez et al. 2005). If the central black hole has itself angular momentum (Genzel et al. 2003b, Aschenbach et al. 2004) Lense-Thirring precession will change the orientation of the angular momentum vector, unless that vector happens to be co-aligned with that of the black hole. We have shown above (3.2) that Lense-Thirring precession is important for the innermost S-stars but is not efficient enough to explain the randomly inclined orbits of all S-cluster stars.

Next, we consider three scenarios from the literature for delivering stars “individually'” (that is, with random, or nearly random orbital properties) to the center: the dissolving sinking cluster (Gerhard 2001, Portegies Zwart, McMillan & Gerhard 2003) bound by an intermediate mass black hole (Hansen & Milosavljevic 2003; Kim, Figer & Morris 2004); the massive binary exchange (Gould & Quillen 2003); and exchange capture with stellar black holes (Alexander & Livio 2004).

The ***dissolving cluster with an intermediate mass black hole scenario*** (Hansen & Milosavljevic 2003) proposes that a dense stellar cluster containing an intermediate mass black hole, perhaps formed by runaway mergers in the cluster core (e.g. Portegies Zwart et al. 2004), has sunk rapidly to the center. The tidal field of the massive black hole dissolved most of the cluster outside the central parsec, but the stars most tightly bound to the intermediate mass black hole were deposited on the ~0.1 pc scale, and subsequently scattered by repeated interactions with the in-spiraling intermediate mass black hole. Hansen & Milosavljevic (2003) hypothesize that the S-stars are in fact O stars from the dissolving cluster that were captured in very bound orbits by a close, nearly disruptive encounter with another star, which stripped their outer envelopes and whittled them down to B-star masses. The general concept of this model received support from recent claims for an intermediate mass black hole embedded in the IRS13 ``cluster'' (Maillard et al. 2004, see however Schödel et al. 2005). However, the weakness of this scenario in explaining the S-stars is that, lacking quantitative calculations, it is unclear whether this method can actually form such a tightly bound cluster, and it is also unclear whether collisional stripping is consistent with the apparent normalcy of the B-stars. Another objection was raised by Kim, Figer & Morris (2004), who simulated this scenario by N-body calculations. They find that in order to carry stars into the inner 0.1 pc, the mass ratio between the intermediate mass black hole and the cluster must be 2 orders of magnitude larger than can feasibly grow in runaway mergers. On the other hand, more massive clusters, which can naturally form a few x $10^3$ $M_\odot$ intermediate mass black hole by core collapse and runaway mergers (Gürkan & Rasio 2004) will leave a tidal tail with many more luminous young stars then are observed in the GC.

The ***massive binary exchange scenario*** (Gould & Quillen 2003) identifies the origin of the B-stars around the massive black hole in an infalling massive cluster. In this scenario the B-star originally had a very massive binary companion (~100 $M_\odot$). The binary presumably originated in a radially infalling, disintegrating massive cluster. The close interaction with the massive black hole led to a 3-body exchange where the B-star switched partners and became bound to the massive black hole. Gould & Quillen (2003) find that the probability for capture in an orbit like that of the star S2 is of the order of a few percent. This mechanism does not explain why it is that only B-stars are captured, and not more massive stars, and neither does it explain the lower bound on the apo-apse found by Ghez et al. (2005). There is also no observational evidence for a significant number of massive O or Wolf-Rayet stars on highly eccentric orbits (Paumard et al. 2005). From a theoretical point of view the weakness of this scenario is the very low joint probability for (i) having a cluster on a radial orbit, (ii) containing a very massive star, (iii) paired in a binary with a much lighter secondary (iv) with the right orbital parameters for 3-body exchange with the massive black hole.

The ***exchange capture with stellar black holes scenario*** (Alexander & Livio 2004) proposes that the B-stars were originally formed far away from the massive black hole where normal star formation is possible (not necessarily in a massive cluster) and were then deflected into eccentric orbits that intersect the dense central concentration of stellar black holes (with masses of 7-10 $M_{\odot}$), which sank to the center from the inner 5 pc due to mass segregation over the lifetime of the Galactic center. This high concentration of the stellar black holes is responsible for the collisional destruction of any tightly bound late type stars at the very center. Once in a while one of these B-stars passes very close to a stellar black hole, knocks it out and replaces it in a bound orbit around the massive black hole (a 3-body exchange involving a massive black hole—stellar black hole "binary'" and the single B-star). This scenario can naturally explain several of the properties of the S-cluster. The central concentration of the S-cluster mirrors that of the highly segregated cluster of stellar black holes. The S-cluster is composed of B-stars because they are most closely matched in mass to the stellar black holes and so have the maximal probability for exchange. The lower bound on the apo-apse, ~0.01 pc (Ghez et al. 2005) corresponds to the point where an exchange requires that the B-star pass so close to the stellar black hole that it is tidally disrupted. The exchange mechanism also predicts a trend toward high eccentricities in the orbits of the captured stars. The recent indication of a high number (>~$10^4$) of neutron stars and stellar-mass black holes in the vicinity of SgrA* (Muno, Pfahl, Baganoff et al. 2004) from Chandra observations supports this scenario. The weakness of this scenario is the high required central concentration of stellar black holes and the large required number of B-stars on eccentric orbits.

All three scenarios have serious weaknesses. The data available at this time cannot decide which, if any, of the models is correct. On balance, 3-body exchange with stellar black holes appear to naturally account for more of the S-stars properties than the other two, but it is unclear whether the necessary conditions hold in the Galactic center. Future SINFONI observations can improve the constraints on such models by exploring the evidence for their required components: an intermediate mass black hole, stellar black holes, dissolving clusters, a reservoir of B-stars in the field and in particular in the disks, and the properties of the stellar mass function of main-sequence stars in the central cusp. More orbital solutions will improve the characterization of the S-stars orbits and may provide additional clues about their origin.

## 4.2 Synchrotron IR flares from accelerated non-thermal electrons

The very steep, 'red' spectral energy distribution of the two small flares we observed with SINFONI strongly argues for a synchrotron emission model with highly energetic, non-thermal electrons. Figure 10 shows the spectral properties of the observed infrared flares superposed on the overall radio to X-ray spectral energy distribution (adopted from Genzel et al. 2003b, see references there to the various data sets). Our observed slope is in good agreement with the synchrotron models of Yuan, Markoff & Falcke (2002), Yuan, Quataert & Narayan (2003, 2004) and Liu, Petrosian & Melia (2004). As predicted by these models, during a flare, perhaps caused by a magnetic field reconnection event, a few percent of the electrons near the event horizon are highly accelerated. Their energy distribution follows a non-thermal power-law for >GeV energies (relativistic $\gamma \geq 10^3$). Our observations thus are in excellent agreement with the prediction that the flaring activity of SgrA* is caused by extra electron acceleration/heating (Markoff et al. 2001). The very steep power-law slope observed for our two smaller flares would suggest that little X-ray synchrotron radiation is emitted. Future observations are required to test whether stronger infrared flares exhibit shallower spectral energy distributions and electron energy distributions, thus whether they are more likely to emit at X-rays as well, or whether the X-ray flares are caused by Compton up-scattering.

## 5. Conclusions

We have presented the first deep, near-diffraction limited infrared integral field spectroscopy of the Galactic Center with an 8m class telescope. Our H/K data with 75mas spatial and 70-100km/s spectral resolution reach a factor ten deeper than previous spectroscopy and give valuable new information on the stars and variable infrared emission from the vicinity of the SgrA* supermassive black hole.

- We find that more than 90% of all K≤16 stars in the stellar cusp within 0.5" of SgrA* are early type stars. Their spectral properties are identical to normal, main sequence B0-B9 stars with moderate (≤150 km/s) rotation.
- From a global fit to the positions and radial velocities of the best six stars within 0.4" of SgrA* we derived their orbital parameters and updated the value for the central mass to be $M=4.06 \times 10^6 (R_o/8\ kpc)^3\ M_\odot$. The orientations of the stellar orbits (angular momenta, apo-apses) appear to be random. The S-stars taken as group are not orbiting in the planes of the two star disks at 0.1-0.5pc of young massive stars. Given their normal stellar properties it is also not very plausible that these stars have formed or been built up in their present location. We favor explanations that have brought these stars to their present location following a strong perturbation of their original orbits.

- The updated estimate for the distance to the Galactic center from the S2 orbit fit is $R_o = 7.62 \pm 0.32$ kpc, resulting in a central mass value of $3.61 \pm 0.32 \times 10^6\ M_\odot$.
- We find the spectral energy distribution of the modest intensity SgrA* flare on July15th to be red ($\alpha' = -4 \pm 1$, $S_\nu \sim \nu^{-\alpha'}$) and featureless. Its slope is consistent with synchrotron models in which the variable SgrA* infrared emission is caused by an increase, due to extra heating/acceleration, of the non-thermal tail of highly relativistic electrons.

***Acknowledgements***. *We are most grateful to the other members of the MPE/ESO/NOVA SINFONI team for their fantastic work on the instrument; it made these observations possible. We are grateful to Margaret Hanson for giving us access to her new spectroscopic atlas of early type stars prior to publication. We thank Fabrice Martins, Sergei Nayakshin, Eliot Quataert and Amiel Sternberg for interesting discussions. T.A. acknowledges support by ISF grant 295/02-1, Minerva grant 8484 and a New Faculty Grant by Sir Djangoly, CBE, of Lindon, UK.. MH is supported by the Euro3D Training Network on Integral Field Spectroscopy, funded by the European Commission under contract No. HPRN-CT-2002-00305.*

## Table 1: Radial velocities of S-Stars

The following table summarizes the radial velocities in km/s (weighted average from the Brγ and HeI absorption line fits) of the S-stars in July and August 2004. The velocities are relative to the local standard of rest. All errors (parentheses, ±) are 1 sigma and include the systematic uncertainties of the wavelength calibration.

| **Star** | **2004.537 (July 2004)** | **2004.632 (August 2004)** |
|---|---|---|
| **S1** | -1060 (30) | -1025 (25) |
| **S2** | -1070 (25) | -1080 (20) |
| **S4** | -540 (35) | -610 (55) |
| **S5** | 75 (90) | -67 (140) |
| **S6** | N/A | 155 (60) |
| **S7** | N/A | -20 (150) |
| **S08** | -450 (70) | -305 (120) |
| **S8** | -15 (45) | -30 (30) |
| **S9** | N/A | 610 (30) |
| **S12** | 210 (50) | 316 (50) |
| **S13** | -230 (70) | -320 (50) |
| **S14** | 300 (100) | 266 (110) |
| **S17** | -1980 (80) | -2120 (80) |
| **id26a (east)** | N/A | -905 (40) |
| **id26b (west)** | N/A | -37 (40) |

## Table 2: Orbital Parameters of S-stars from global fit[a]

Global fit parameters for $R_0$=8.0 kpc

Offset center of mass from nominal SgrA* position:

RA=0.0018 (0.0012) (arcsec)

DEC =-0.0051 (0.0012) (arcsec)

Central mass: $M_0$=4.06 (0.38)x$10^6$ $M_\odot$

Overall $\chi^2$ is 295 for 224 degrees of freedom, or $\chi^2_{(r)}$=1.3

| parameter[b]/star | S1 | S2 | S8 | S12 | S13 | S14 |
|---|---|---|---|---|---|---|
| **semi-major axis *a* (arcsec)** | 0.412 (0.024) | 0.1226 (0.0025) | 0.329 (0.018) | 0.286 (0.012) | 0.219 (0.058) | 0.225 (0.022) |
| **numerical eccentricity *e*** | 0.358 (0.036) | 0.8760 (0.0072) | 0.927 (0.019) | 0.9020 (0.0047) | 0.395 (0.032) | 0.9389 (0.0078) |
| **orbital period *P* (years)** | 94.1 (9.0) | 15.24 (0.36) | 67.2 (5.5) | 54.4 (3.5) | 36 (15) | 38.0 (5.7) |
| **epoch of peri-astron passage $t_0$** | 2002.6 (0.6) | 2002.315 (0.012) | 1987.71 (0.81) | 1995.628 (0.016) | 2006.1 (1.4) | 2000.156 (0.052) |
| **inclination *i* (degrees)** | 120.5 (1.0) | 131.9 (1.3) | 60.6 (5.3) | 32.8 (1.6) | 11 (35) | 97.3 (2.2) |
| **position angle of the ascending node Ω (degrees)** | 341.5 (0.9) | 221.9 (1.3) | 141.4 (1.9) | 233.3 (4.6) | 100 (198) | 228.5 (1.7) |
| **longitude of periastron *ω* (degrees)** | 129.8 (4.7) | 62.6 (1.4) | 159.2 (1.8) | 311.8 (3.6) | 250 (161) | 344.7 (2.2) |
| **period of Lense-Thirring precession (yrs)[c]** | 1.74e9 (0.60e9) | 6.3e6 (2.2e6) | 5.7e7 (2.1e7) | 5.8e7 (2.0e7) | 2.48e8 (0.86e8) | 1.42e7 (0.51e6) |

[a] errors (parentheses) are 1σ and ± and include an uncertainty of ±0.32 kpc in the distance to the Galactic center (section 3.3).

[b] definition of orbital elements according to Aller et al. (1982) in Landolt-Börnstein numerical data and relationships in science and technology: $a$ = semi-axis major; $P$ = orbital period (period of revolution); $e$ = numerical eccentricity; $t_0$ = epoch of periastron passage (corresponds to minimum distance from central black hole); $\Omega$ = position angle of ascending node for equinox 2000. The nodes are the points of intersection of the relative orbit with the plane tangential to the celestial sphere at the position of the central black hole. The star recedes from us at the ascending node; $i$ = inclination of the ascending node = angle between the orbital plane and the plane tangential to the celestial sphere. For direct (counter-clockwise) apparent motion of the star, it is given between 0° and 90°, for retrograde motion between 90° and 180°. This angle is measured at the ascending node, from the direction of increasing position angle (in the apparent plane) to the direction of the motion (in the true orbital plane); $\omega$ = longitude of periastron = the angle between the radius vector of the ascending node and that of the periastron in the true orbit, counted form the node in the direction of the orbital motion.

[c] from equation 5 of Levin & Beloborodov (2003), with spin parameter a=0.52+/-0.15 (Genzel et al. 2003b)

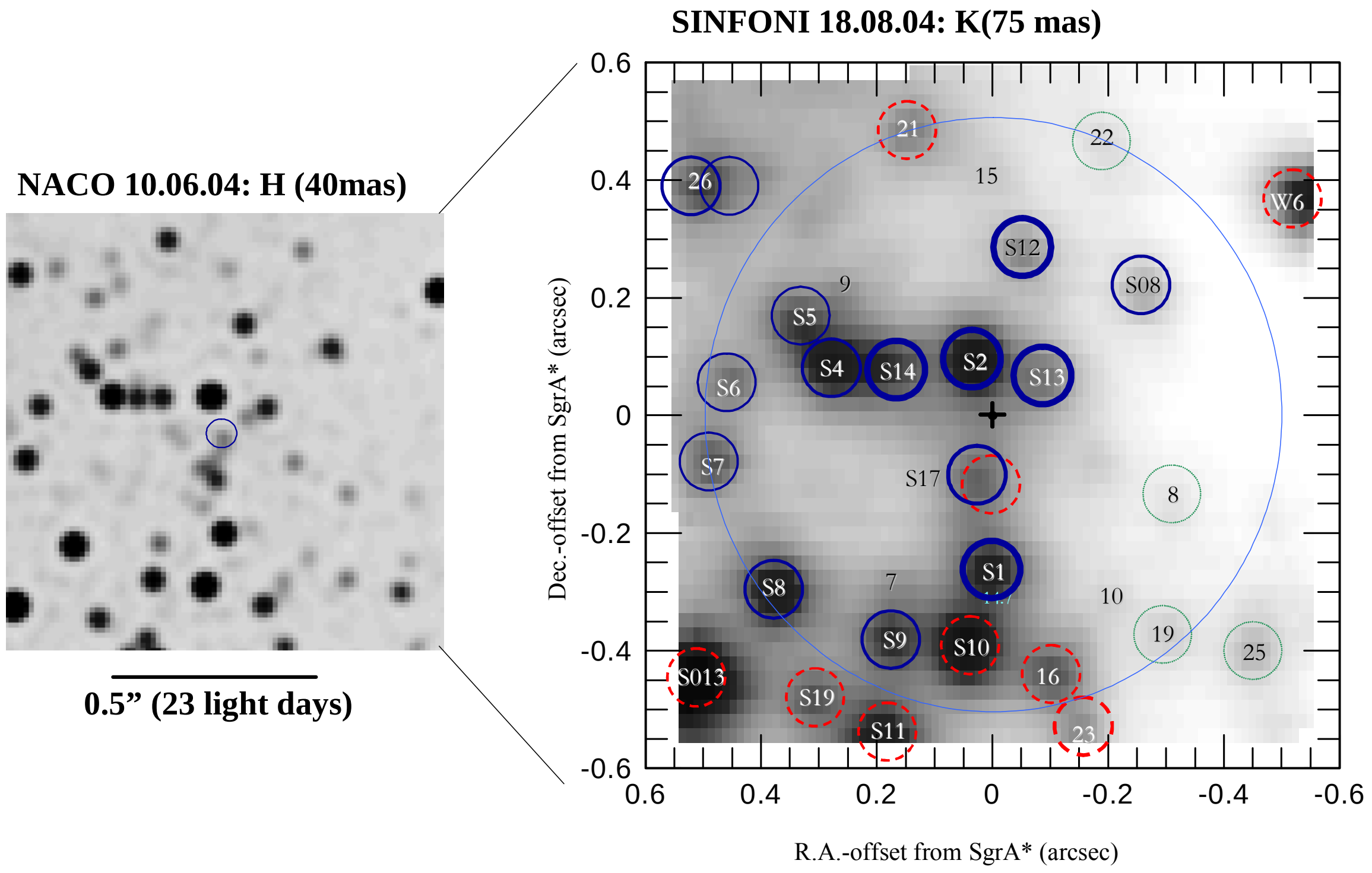

**Figure 1.** NACO and SINFONI H/K-images of the central arcsecond. The left inset shows a NACO H-band image (Lucy cleaned and reconvolved to 40mas resolution) taken on June 10th, 2004. The color scale is logarithmic. The magnitude difference between the faintest and brightest stars in the image is about 4.4 mag and the faintest stars in the image have K~17.8. The position of the central black hole and radio source SgrA* is indicated by a circle. The right inset is a K-band image constructed from all of the SINFONI data from August 18/19th, 2004 by collapsing the data cube in spectral dimension (grey scale is logarithmic). The spatial resolution of the SINFONI images is 75mas FWHM and the Strehl ratio is about 10%. The adaptive optics module was locked on a V=14.6mag star ~20” NE of SgrA*. All K<17 stars visible in the SINFONI image (and the NACO image) are marked by a circle, along with their identification (Schödel et al. 2003). The form and thickness of each circle encodes the spectral identification (continuous: early type, dashed: late type, thin dotted: uncertain but not late type) and proper motion velocity. Early and late type stars with proper motions ≥1000 km/s are marked by thick circles, stars with proper motions between 500 and 1000 km/s by medium thick circles and stars with proper motions <500 km/s by thin circles.

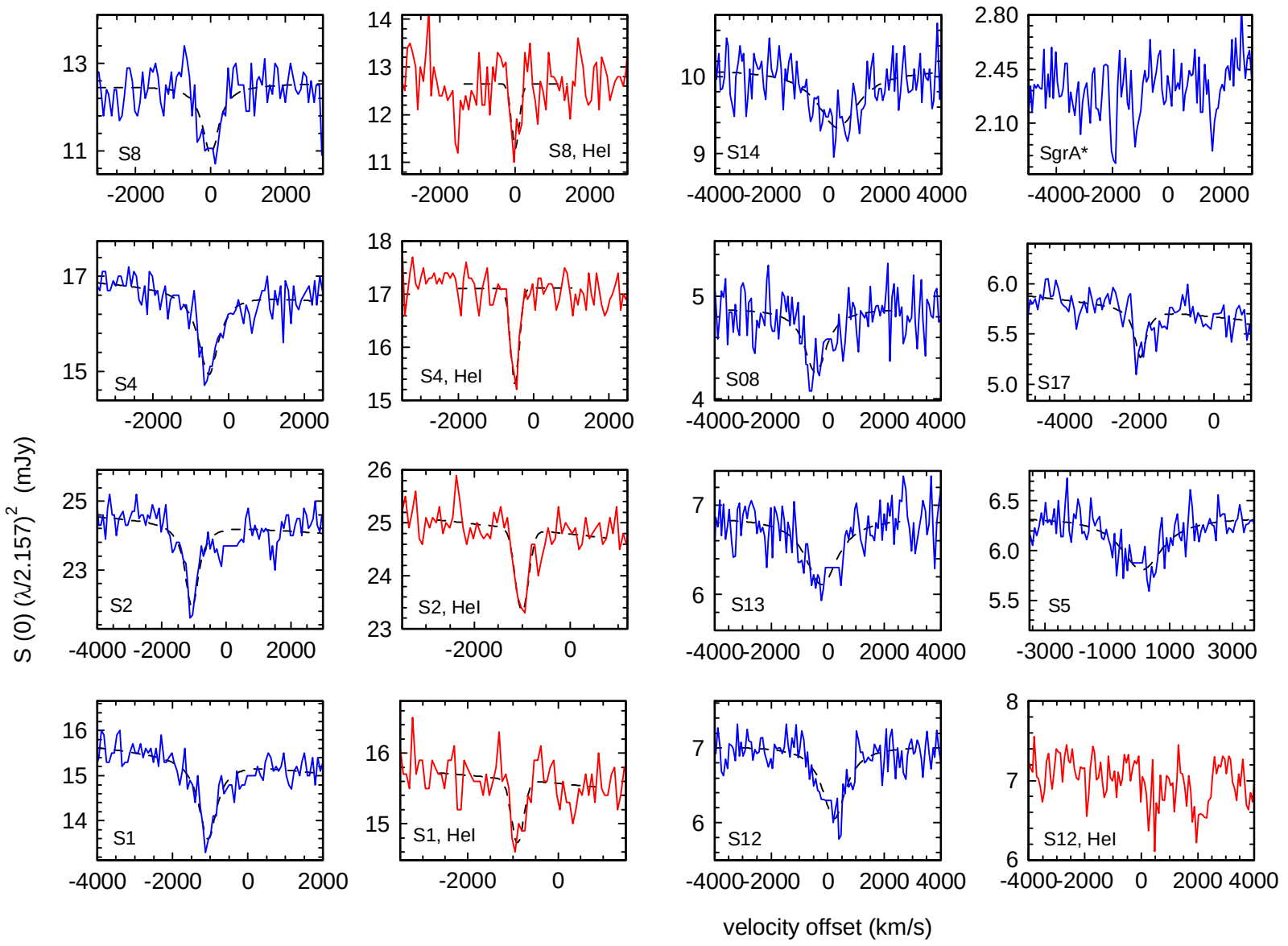

**Figure 2.** Extracted spectra near the 2.1661µm HI Brγ (7-4) line and the 2.1127µm HeI line (labeled) for 10 of the 11 K<16 mag stars in the central ~0.5", plus a spectrum toward SgrA* itself (upper right inset), from the H+K (R~3000) data set of July 15th. In all cases we extracted spectra ~75mas diameter circular apertures centered on the star and subtracted the average of several nearby background apertures that are not contaminated by strong stellar emission. To eliminate any residual weak interstellar Brγ emission we interpolated some of the spectra in the region -100 to +200 km/s LSR region when necessary. Vertical axis are dereddened (A(2.157µm)=2.8mag) flux densities, multiplied by $(\lambda/2.157\mu m)^2$ to approximately level the spectra. Dashed curves are the best Lorentzian (Brγ) and Gaussian (HeI) fits (see Table 1).

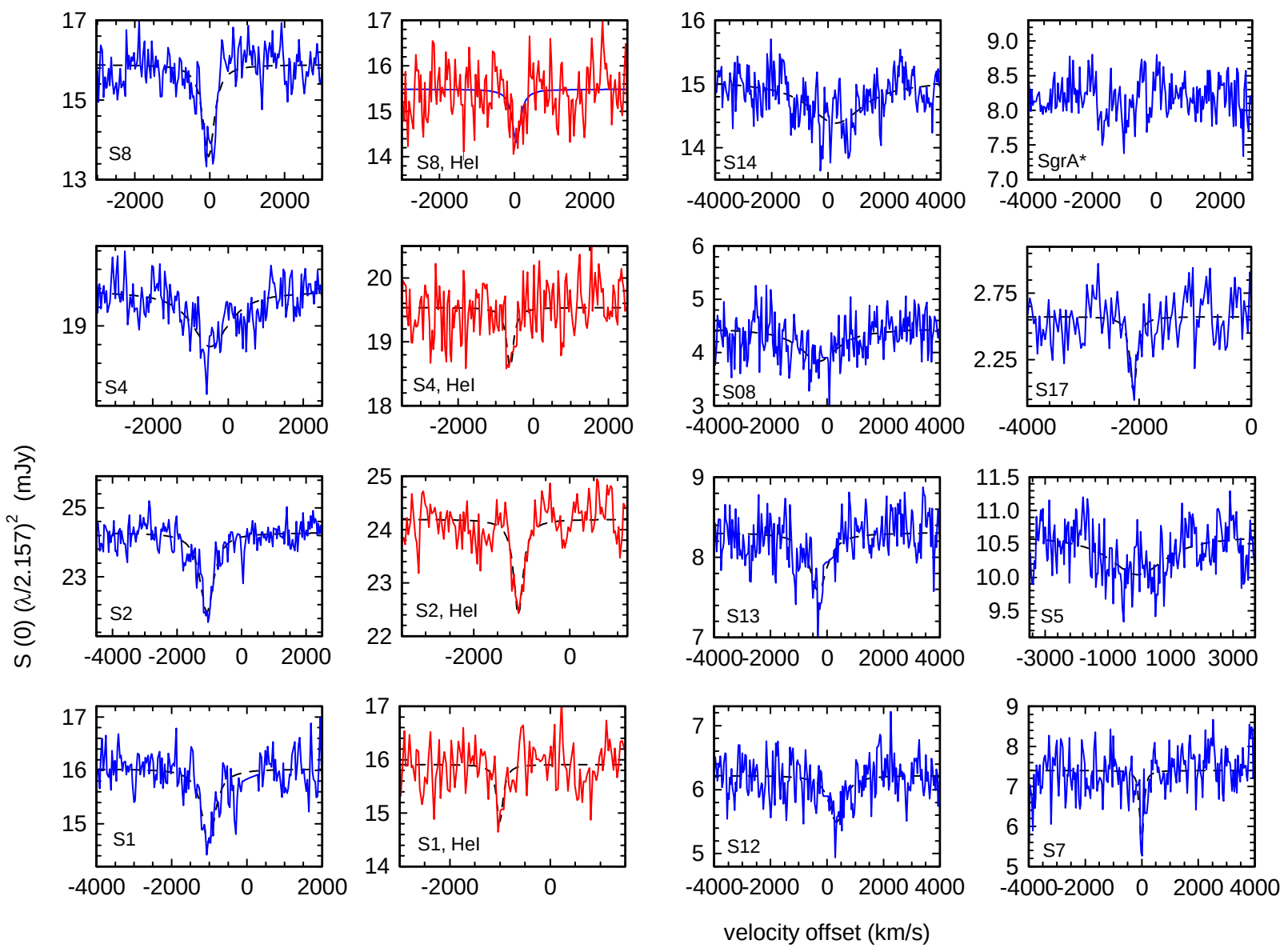

**Figure 3.** Same as in Figure 2 but for the August K-band data set (R~4400). Since the integration time varied significantly over the field, the spectra of some of the stars away from the center are relatively noisier than in Figure 2.

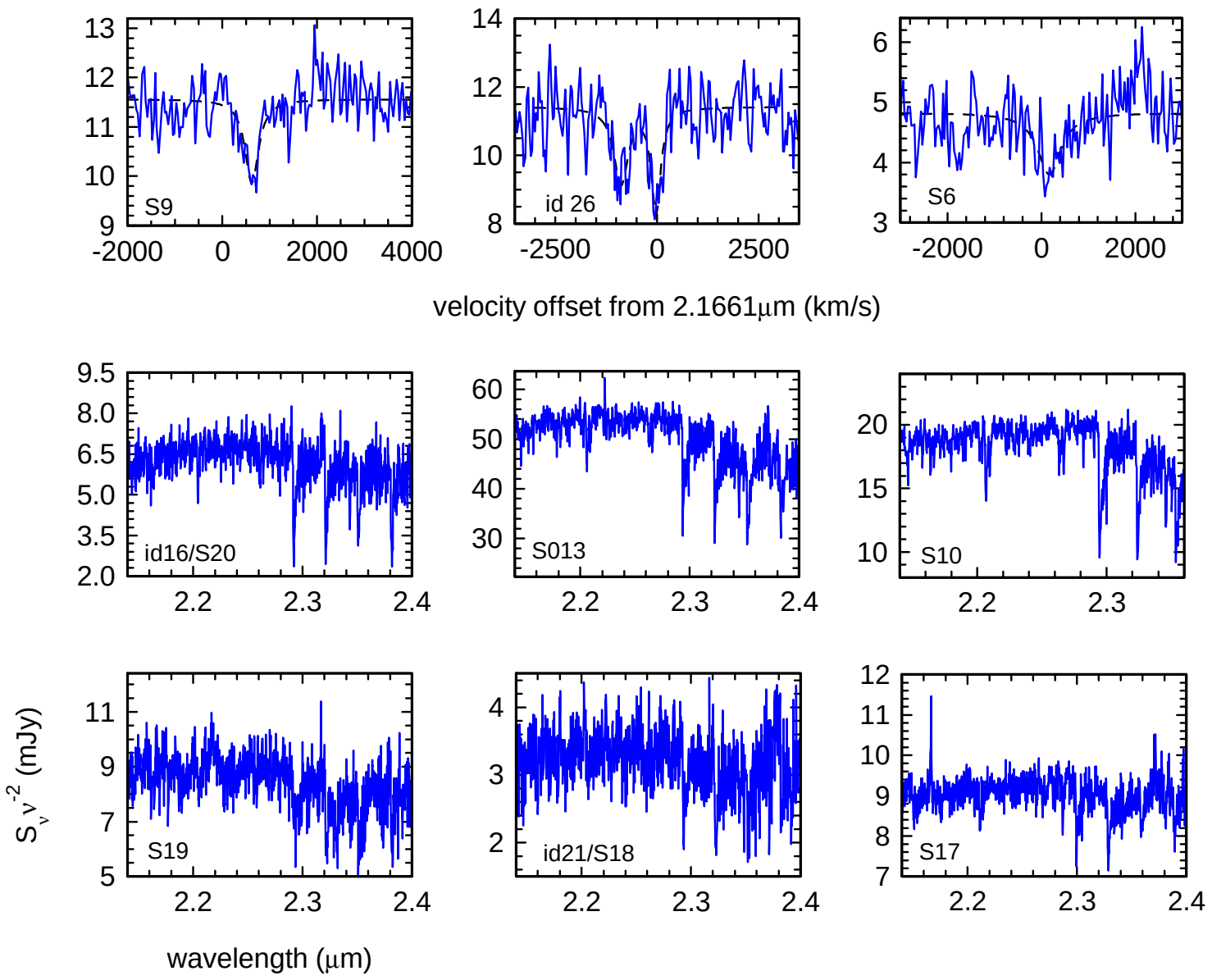

**Figure 4.** Same as in Figure 2 but for the August K-band data set (R~4400). Since the integration time varies significantly over the field, some of the stars away from the center are noisier than in Figure 2. For the star 'id26' (Schödel et al. 2003) there appear to be two Brγ absorption features. Inspection of the data cube shows that these features originate from two nearby stars that are well resolved on the H-band NACO image (see Figure 1) but poorly resolved in the SINFONI data set. In addition to the 15 early type stars in this Figure and Figures 3 and 2 we also show here six of the 9 late type stars in the SINFONI field of view.

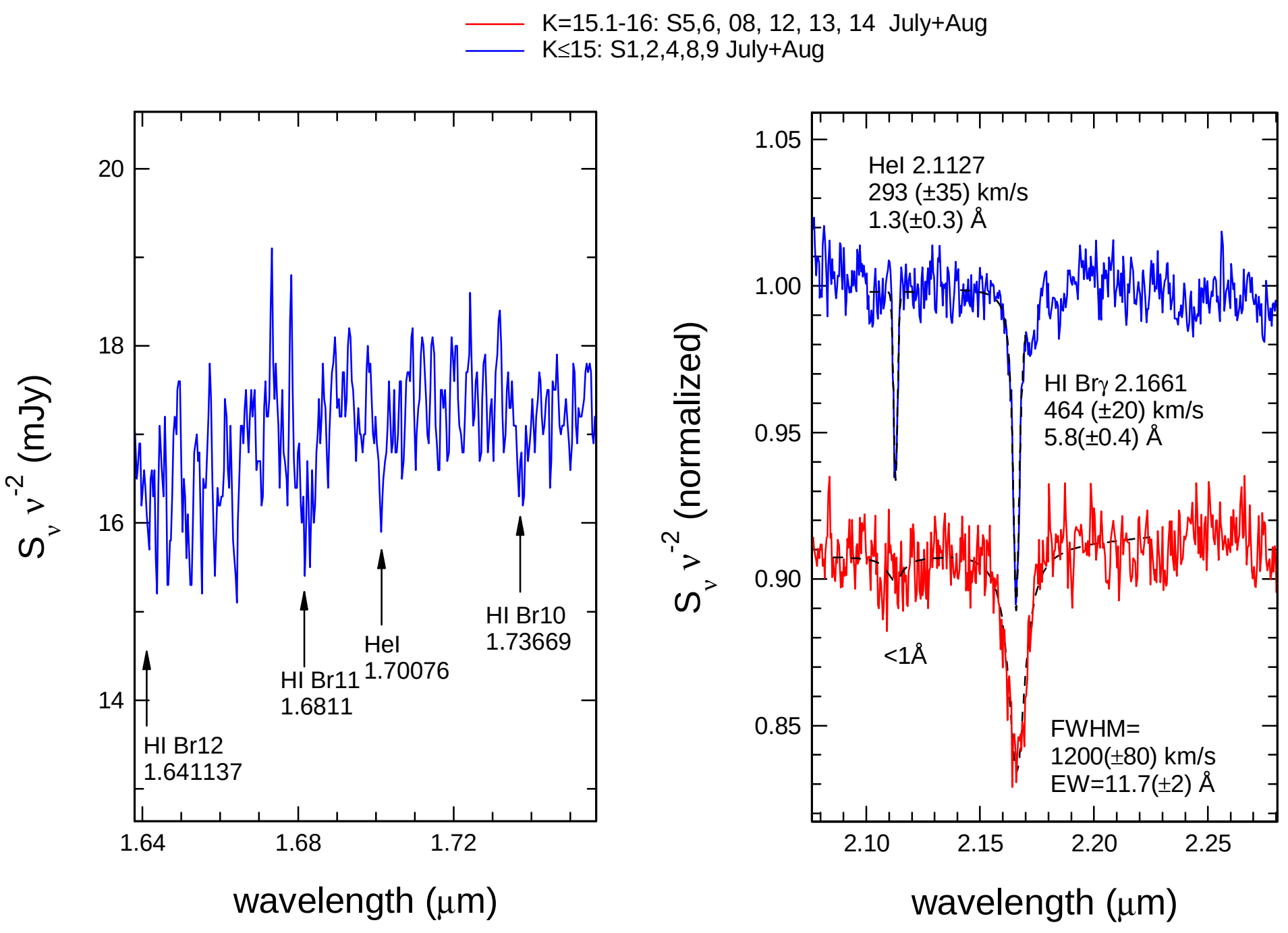

**Figure 5.** Co-added H- and K-band spectra of the S-stars split into a group of brighter stars (right top and left: K=13.9 to 15: S1, S2, S4, S8, S9) and a group of fainter stars (right bottom: K=15.2 to 15.8: S5, S08, S12, S13, S14). We used both the July and August data sets, interpolated to the same R~3000 spectral resolution. Before co-adding we shifted each spectrum in velocity to the peak of its absorption. The vertical axis is dereddened, normalized flux density (A(2.157μm)=2.8mag) multiplied by $(\lambda/2.157\mu m)^2$ to approximately take out the Rayleigh-Jeans slope of a hot blackbody. The two groups of S-stars exhibit significantly different spectral properties.

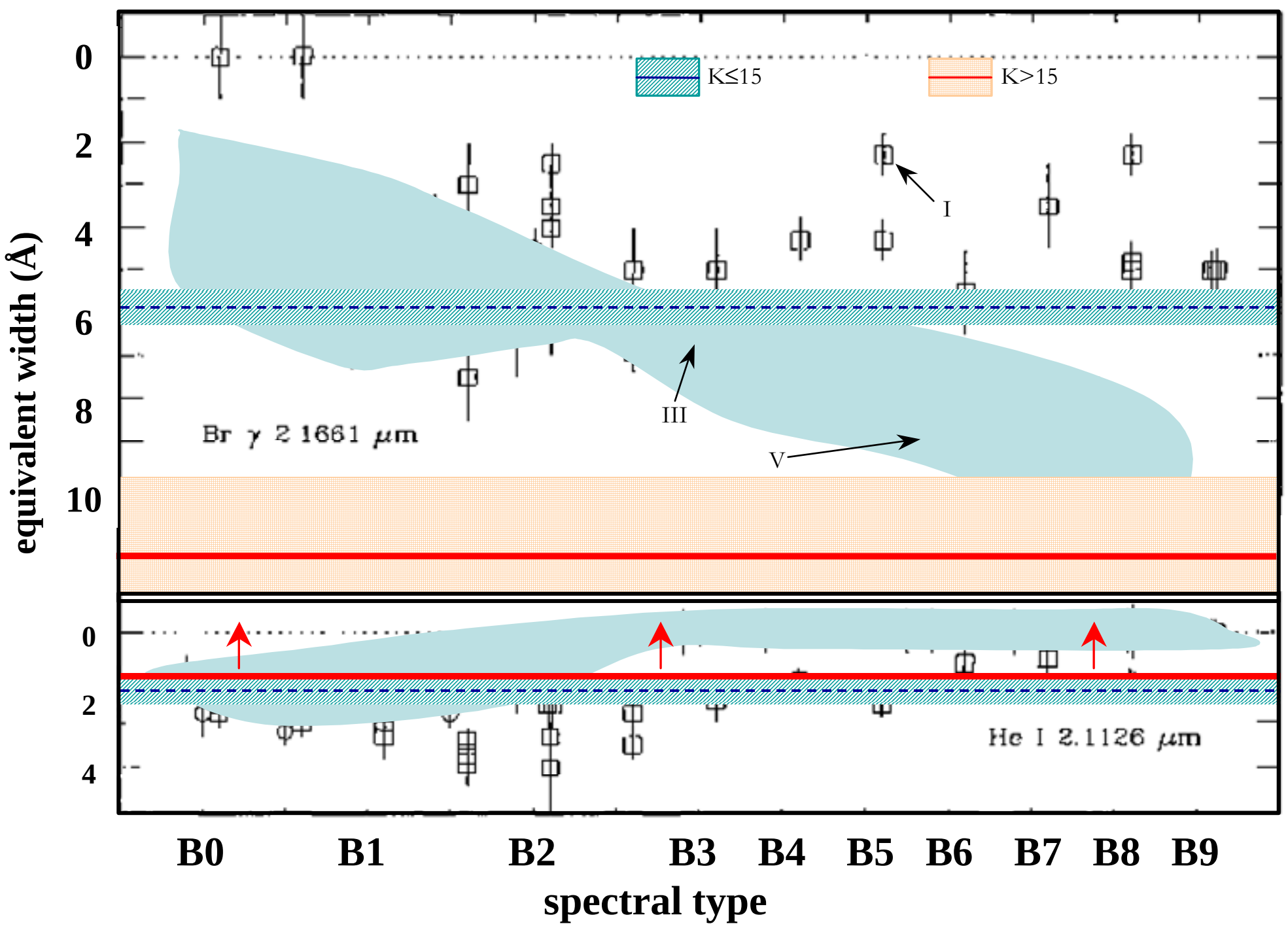

**Figure 6.** Equivalent width of HI Brγ (top panel) and 2.1127µm HeI (bottom panel) for main sequence (crosses, connected by shaded region), giant (circles) and super-giant (squares) B-stars, from the compilation of solar neighborhood stars by Hanson et al. (1996). The horizontal dashed line with diagonally hatched, 1σ error regions marks the constraints for the bright group Galactic Center S-stars (K≤15). The thick continuous line and cross-hatched error region mark the constraints for the fainter group (15<K<15.8) of S-stars. Adapted from Hanson et al. (1996).

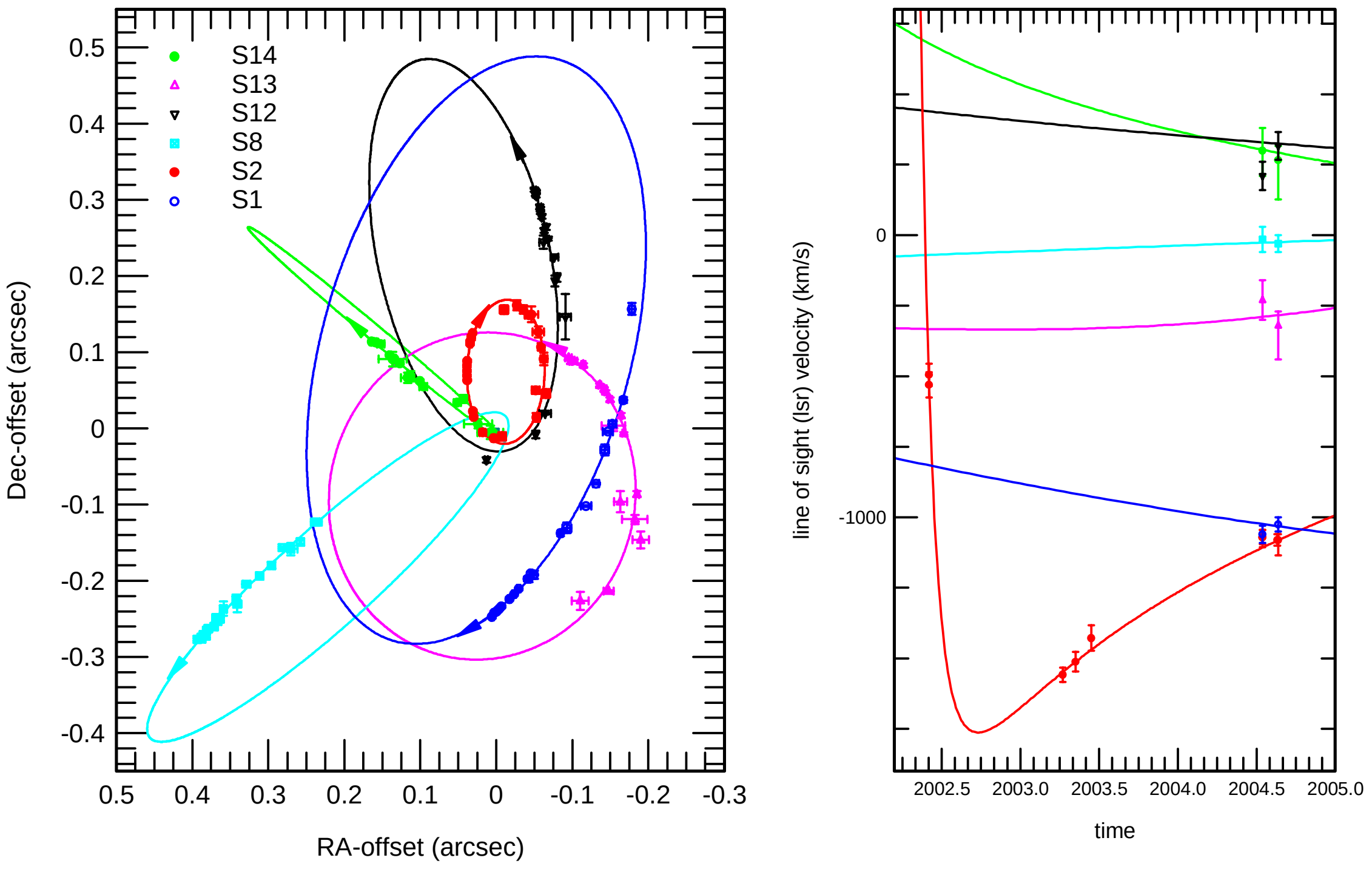

**Figure 7.** Projection on the sky (left inset) and in time/radial velocity (right inset) of the 6 S-stars included in the fitting (see also Schödel et al. 2003). The various color curves are the result of the best global fit to the spatial and radial velocity data of S1, S2, S8, S12, S13 and S14. The orbital parameters are listed in Table 2. The assumed distance is 8 kpc.

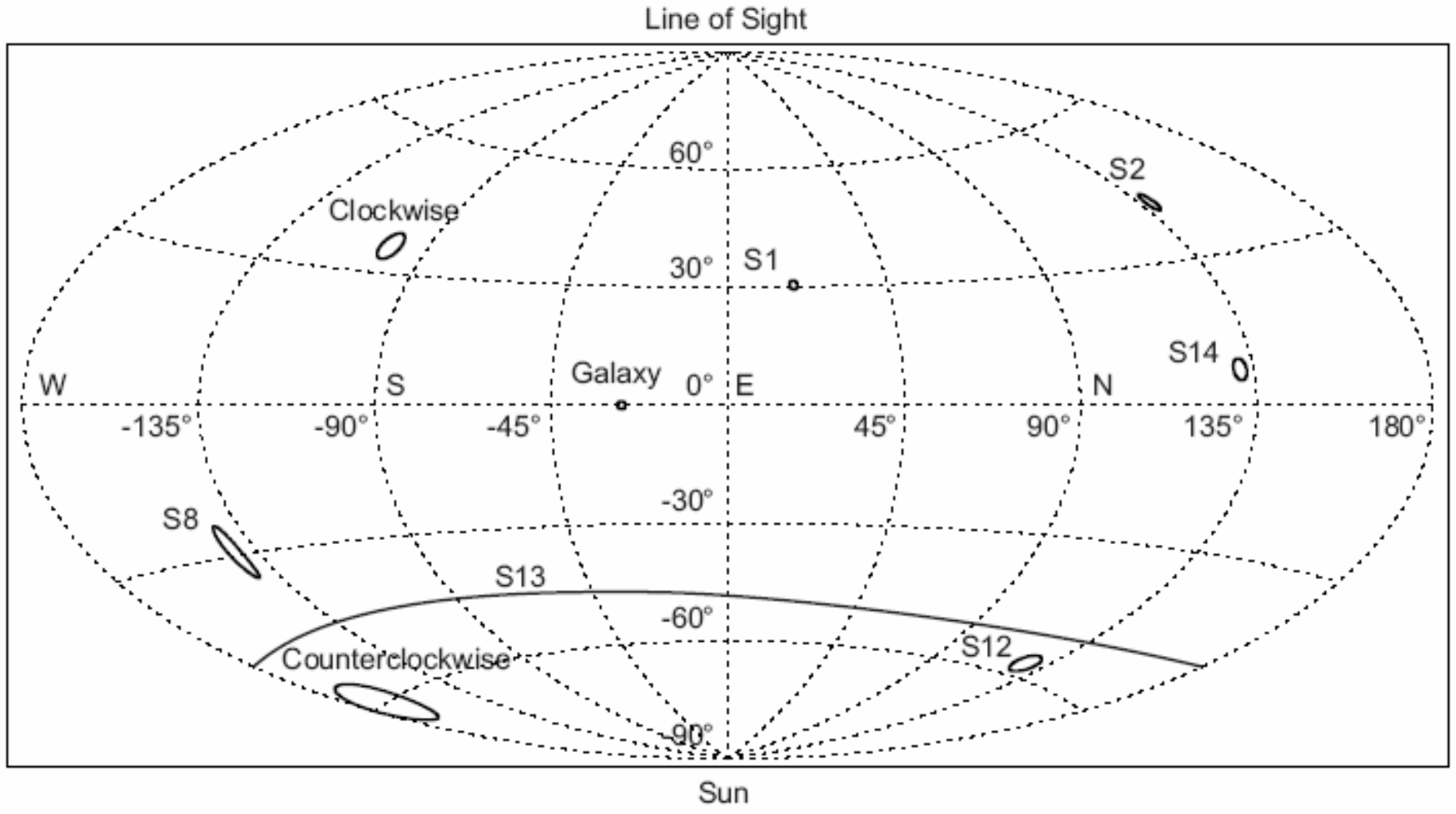

**Figure 8a.** Directions of angular momentum vector on the sky of the six best determined orbits of Fig.7, along with the orientations of the angular momentum vector of the Galaxy and of the two young star disks at p~2-8" from SgrA* (Paumard et al. 2005, see also Levin & Boloborodov 2003, Genzel et al. 2003a) . The diagram shows the one-sigma error ellipses for the direction of the angular momentum using an Aitoff map projection. The directions of angular momentum appear random, and - with the exception of S13 with its comparably large error - all measured directions of angular momentum are different from the rotation axes of the counter-rotating stellar discs at larger radii and the Galaxy. The star S13 could be – but not necessarily is – part of the counterclockwise system.

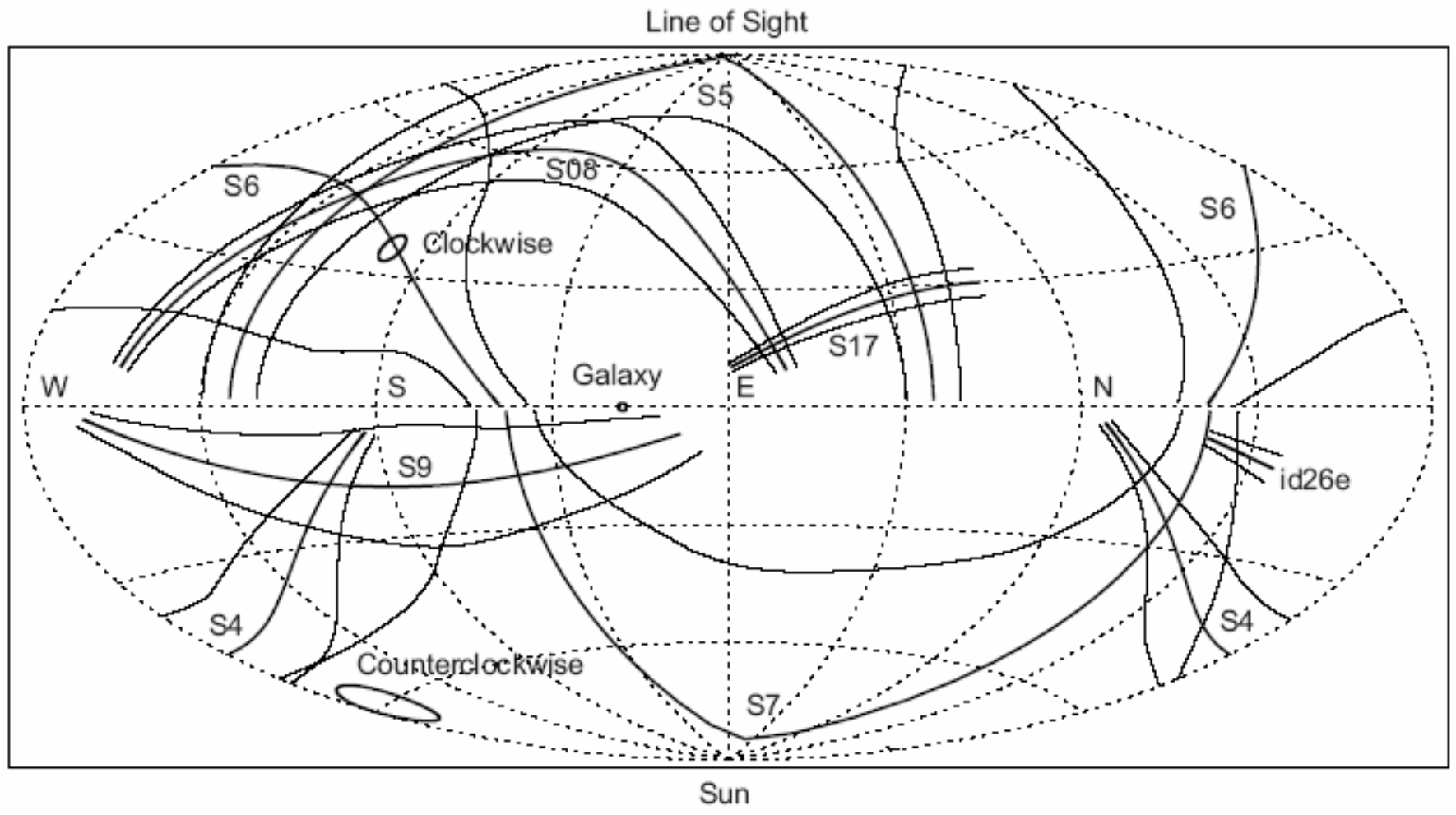

**Figure 8b.** Direction of angular momentum of the eight stars for which no full three-dimensional orbital solution is available, but for which we could measure the position, proper motion, and radial velocities. One parameter – the position of the star along the line of sight – is missing for determining the direction of the angular momentum. We thus show lines of directions of the angular momentum for a range of possible positions along the line of sight (thick lines) and the one-sigma error range (thin lines). We limited the range of positions along the line of sight assuming that the star is gravitationally bound to the central black hole (for a distance of 8.0 kpc and a central mass of 4.06 million solar masses), i.e. that the measured velocity is always smaller than the escape velocity for given position along the line of sight. The angular momentum of six (out of eight) stars (S4, S5, S08, S9, S17, id26e) is not consistent with a motion in the plane of the counter-rotating stellar discs (see Figure 8a) from Genzel et al. (2003a, 2005) and Levin &

Boloborodov (2003). Two stars (S6, S7) could be – but not necessarily are - part of the clockwise system, the counter-clockwise system, respectively.

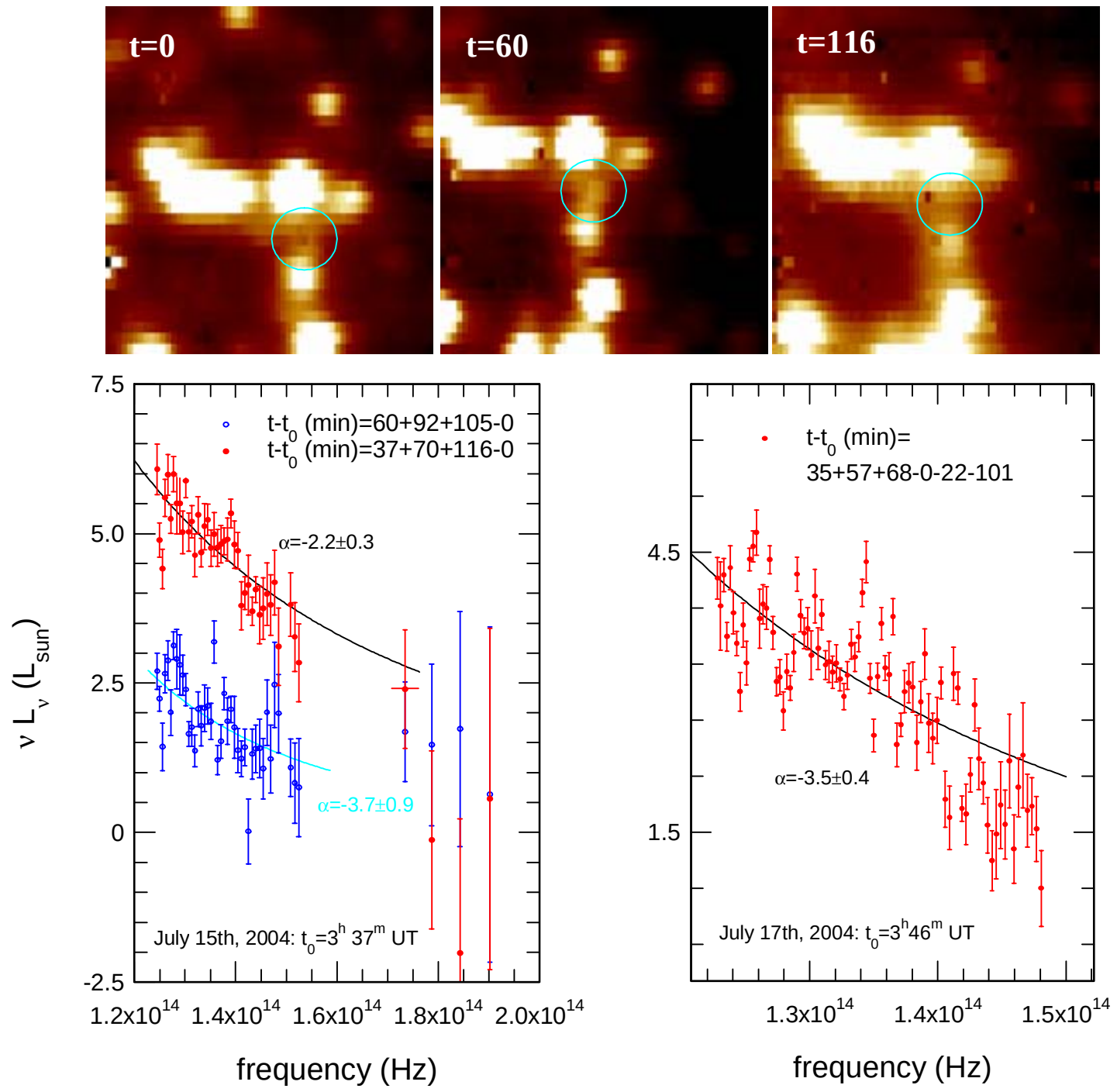

**Figure 9.** Observations of infrared flares from SgrA* with SINFONI. Top: Wavelength-collapsed data cubes at three different times on July 15$^{th}$, 2004: 'before' (t=0), 'peak' (t=60min) and 'after' (t=116min) (relative to $t_o=3^h\ 37^m$ UT). Bottom left: dereddened $\nu L_\nu$ spectral energy distribution of the SgrA* flare on July 15$^{th}$, 2003. The filled circles ( plus 1σ error bars) shows the H/K spectral energy distribution averaged over three 10 minute frames near the flare maximum, while the open circles are averaged over three frames at the rising and falling flanks of the flare. In both cases we calibrated the wavelength dependent Strehl ratio by dividing the observed data by S2. To correct for scattered light/spillover and foreground/background stellar emission along the line of sight to SgrA* we also subtracted the 'before' spectrum from the flare spectra. The spectral data are binned to $\Delta\lambda$=0.0105μm above 1.97μm, and 0.05μm below. Solid curves mark the best power law fits with spectral index (and 1σ errors) given in the figure ($\nu L_\nu \sim \nu^\alpha$).

Bottom right: Same but for the flare on July 17th, 2004. Here we show the K-band spectral energy distribution of the average of three 10 minutes integrations near the peak of the flare (minus the average of three ‘off’ frames). The spectral binning width was 0.0049μm.

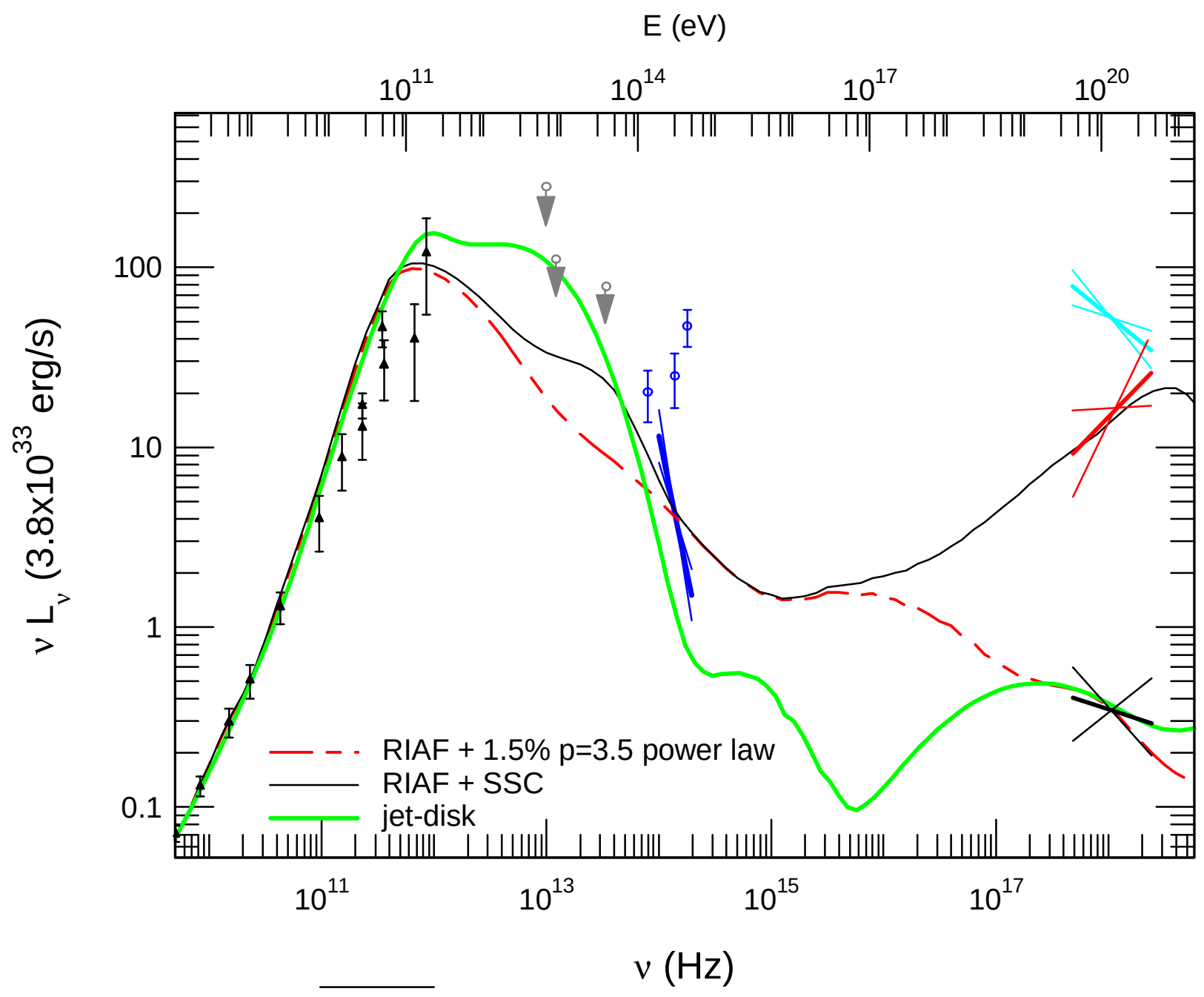

**Figure 10.** Spectral energy distribution (SED) of SgrA* from radio to X-rays (adopted from Genzel et al. 2003b), plus the July15th and July17th 2004 flare SEDs obtained from our new SINFONI data (blue lines with ±2σ error regions). The wavelength range of the SINFONI data is 1.7 to 2.4μm and the wavelength extent has been exaggerated in the figure for clarity. We also plot the various extinction and absorption corrected radio to X-ray band luminosities $\nu L_\nu$ (energy emitted per logarithmic energy interval for an assumed distance of 8 kpc) as a function of frequency, or energy. All error bars are ±1σ. Black triangles denote the quiescent radio spectrum of SgrA*. Open grey circles denote various IR upper limits from the literature. The three X-rays data ranges are (from bottom to top) the quiescent state as determined with Chandra (black), the fall 2000 Chandra flare (red) and the fall 2002 XMM flare 21 (light blue). Open red squares with crosses mark the peak emission (minus quiescent emission) observed in the four 2003 VLT flares (Genzel et al. 2003b). Open blue circles denote the dereddened H, $K_s$ and L' luminosities of the quiescent state, derived from the local background subtracted flux density of the point source at the position of SgrA*, thus eliminating the contribution from any extended diffuse light from the stellar cusp around the black hole. The thick green solid curve is the jet-starved disk model of reference (Yuan et al. 2002). The red long dash-short dash curve is a radiatively inefficient accretion flow (RIAF) model of the quiescent emission where in addition to the thermal electron population of (Yuan et al. 2004) 1.5% of the electrons are in a non-thermal power law energy spectrum of exponent p=-3.5. The black thin solid curve is a RIAF model of the flares with 5.5% of the electrons in a power law of p=-1 (Yuan et al. 2004). The long dash thick curve is a RIAF flare model of the flares with a synchrotron-self Compton component (Yuan et al. 2004).